\documentclass[amstex,twocolumn,showpacs,floats,floatfix,superscriptaddress,aps,pra]{revtex4}
\usepackage{amssymb}
\usepackage{amsmath}
\usepackage{calc}
\usepackage{graphicx}
\usepackage{bm}

\def\be{ \begin{equation} }
\def\ee{ \end{equation} }
\def\bea{ \begin{eqnarray} }
\def\eea{ \end{eqnarray} }
\def\bse{ \begin{subequations} }
\def\ese{ \end{subequations} }

\def\i{\,\text{i}}
\def\e{\,\text{e}}
\def\d{\,\text{d}}

\def\Red{} \def\red{\Red}
\def\Black{} \def\black{\Black}

\begin{document}

\author{G. S. Vasilev}
\affiliation{Department of Physics, University of Oxford, Parks Road, OX1 3PU Oxford, United Kingdom}
\affiliation{Department of Physics, Sofia University, James Bourchier 5 blvd, 1164 Sofia, Bulgaria}
\author{A. Kuhn}
\affiliation{Department of Physics, University of Oxford, Parks Road, OX1 3PU Oxford, United Kingdom}
\author{N. V. Vitanov}
\affiliation{Department of Physics, Sofia University, James Bourchier 5 blvd, 1164 Sofia, Bulgaria}
\affiliation{Institute of Solid State Physics, Bulgarian Academy of Sciences, Tsarigradsko chauss\'{e}e 72, 1784 Sofia, Bulgaria}
\title{Optimum pulse shapes for stimulated Raman adiabatic passage}
\date{\today }

\begin{abstract}
Stimulated Raman adiabatic passage (STIRAP), driven with pulses of
optimum shape and delay has the potential of reaching fidelities
high enough to make it suitable for fault-tolerant quantum
information processing. The optimum pulse shapes are obtained upon
reduction of STIRAP to effective two-state systems. We use the
Dykhne-Davis-Pechukas (DDP) method to minimize nonadiabatic
transitions and to maximize the fidelity of STIRAP. This results
in a particular relation between the pulse shapes of the two
fields driving the Raman process. The DDP-optimized version of
STIRAP maintains its robustness against variations in the pulse
intensities and durations, the single-photon detuning and possible
losses from the intermediate state.
\end{abstract}

\maketitle

\section{Introduction}

Stimulated Raman adiabatic passage (STIRAP) is a well established
and widely used technique for coherent population transfer in
atoms and molecules \cite{STIRAP}. STIRAP uses two delayed but
partially overlapping laser pulses, pump and Stokes, which drive a
three-state $\Lambda $-system $\psi_1\rightarrow \psi_2\rightarrow
\psi_3$. The STIRAP technique transfers the population
adiabatically from the initially populated state $\psi_1$ to the
target state $\psi_3$. If the pulses are ordered
counterintuitively, i.e. the Stokes pulse precedes the pump pulse,
two-photon resonance is maintained, and adiabatic evolution is
enforced, then complete population transfer from $\psi_1$ to
$\psi_3$ occurs. Throughout this process, no population is placed
in the (possibly lossy) intermediate state $\psi_2$. Various
aspects of STIRAP have been subjects of intense research, both
theoretically and experimentally \cite{STIRAP-reviews}.

Because STIRAP is an adiabatic technique it is insensitive to
small to moderate variations in most of the experimental
parameters, such as pulse amplitudes, widths,  delay, and
single-photon detuning. A particularly remarkable and very useful
feature of STIRAP is its insensitivity to the properties of the
intermediate state $\psi_2$. For instance, STIRAP has been
demonstrated with high efficiency even for interaction durations
exceeding the lifetime of $\psi_2$ by a factor of 100
\cite{STIRAP-reviews}. For these reasons STIRAP is a very
attractive technique for quantum information processing (QIP)
\cite{STIRAP-MPQ,STIRAP-QIP}. However, it is widely recognized
that QIP requires very high fidelities, with the admissible error
of gate operations being below $10^{-4}$ for a reliable quantum
processor \cite{Nielsen00,fault tolerant QC}. Such an extreme
accuracy has not yet been demonstrated for STIRAP, as an accuracy
of 90-95\% was sufficient for most traditional applications. When
trying to increase the fidelity beyond this number one faces
various obstacles related mainly to nonadiabatic transitions.
Being an adiabatic technique, STIRAP reaches an efficiency of
unity only in the adiabatic limit; however, the latter is
approached only asymptotically as the pulse areas increase. For
QIP, the pulse areas needed are so large that they may violate
various restrictions of a real experiment.

In this paper we propose how to achieve an ultrahigh fidelity in
STIRAP, and thus make it fully suitable for QIP by suitably shaped
pulses. We utilize a recent idea of Guerin \emph{et al.}
\cite{Guerin} who applied the well-known Dykhne-Davis-Pechukas
(DDP) method \cite{Davis76} to optimize adiabatic passage in a
two-state system. In order to adapt this approach to STIRAP, we
reduce the three-level Raman system to effective two-state systems
in two limits: on exact resonance and for large single-photon
detuning. The optimization, which minimizes the nonadiabatic
transitions and maximizes the fidelity, leads to a particular
relation between the pulse shapes of the driving pump and Stokes
fields.

It should be noted that a fidelity of unity can also be achieved
by a resonant $\pi $-pulse in a two-state transition. However,
resonant techniques suffer from their sensitivity to parameter
fluctuations. The optimized version of STIRAP presented here
features both a very high fidelity and a robustness against
variations in the intensities and the single-photon detuning.

This paper is organized as follows: In section
\ref{Sec-background} we review the DDP method and the optimization
of two-state adiabatic passage. Then we extend this idea to STIRAP
in section \ref{Sec-STIRAP} and discuss examples in section
\ref{Sec-implementation}. In section \ref{Sec-f-STIRAP} we extend
these ideas to fractional STIRAP (f-STIRAP), which creates a
coherent superposition of $\psi_1$ and $\psi_3$. We summarize the
results in the concluding section.

\section{Optimization of adiabatic passage between two states \label{Sec-background}}

\subsection{Dykhne-Davis-Pechukas (DDP) approximation}

The probability amplitudes in a two-state system $\mathbf{a}(t)=\left[a_1(t),a_2(t)\right]^{T}$ satisfy the Schr\"{o}dinger equation,
\be
\text{i}\hbar \frac{\text{d}}{\text{d}t}\mathbf{a}(t)=\mathbf{H}(t)\mathbf{a}(t),  \label{Schrodinger-2SS}
\ee
where the Hamiltonian in the rotating-wave approximation (RWA) reads \cite{B.Shore}
\be
\mathbf{H}(t) = \tfrac12\hbar \left[ \begin{array}{cc} 0 & \Omega (t) \\ \Omega (t) & 2\Delta (t) \end{array} \right] .  \label{H2}
\ee
The detuning $\Delta =\omega_0-\omega$ is the difference between the transition frequency $\omega_0$ and the carrier laser frequency $\omega$.
The time-varying Rabi frequency $\Omega (t) = \left\vert dE(t)\right\vert/\hbar$ describes the laser-atom interaction, where $d$ is the electric dipole moment for the $\psi_1\leftrightarrow \psi_2$ transition
 and $E(t)$ is the laser electric field envelope.

A very accurate technique for deriving the transition probability in the near-adiabatic regime is the Dykhne-Davis-Pechukas (DDP) approximation \cite{Davis76}.
The DDP formula gives the following expression for the probability for nonadiabatic transitions
\be
P\approx e^{-2\text{Im}D(t_0)},  \label{DP-1}
\ee
where
\be
D(t_0)=\int_0^{t_0}\varepsilon(t)dt  \label{D(Tc)}
\ee
is an integral over the splitting
$\varepsilon(t)=\sqrt{\Omega (t)^2+\Delta (t)^2}$ of the
eigenenergies of the Hamiltonian (\ref{H2}). The point $t_0$ (the
transition point) is defined as the (complex) zero of the
quasienergy
splitting, $\varepsilon(t_0)=0$, which lies in the upper half of the complex $t$%
-plane (i.e., with Im$\mathrm{\,}t_0>0$). Equation (\ref{DP-1})
gives the correct asymptotic probability for nonadiabatic
transitions provided:
 (i) the quasienergy splitting $\varepsilon(t)$ does not vanish for real $t$, including $\pm \infty $;
 (ii) $\varepsilon(t)$ is analytic and single-valued at least throughout a region of the complex $t$-plane that includes the region from the real axis to the transition point $t_0$;
 (iii) the transition point $t_0$ is well separated from the other quasienergy zero points (if any), and from possible singularities;
 (iv) there exists a level (or Stokes) line defined by \text{Im}$D(t)=\text{Im} D(t_0)$, which extends from $-\infty$ to $+\infty$ and passes through $t_0$.

For the case of multiple zero points in the upper $t$-plane, Eq.~(\ref{DP-1}) can be generalized to include the contributions from all these $N$ zero points $t_{k}$ as
\be
P\approx \left\vert \sum\nolimits_{k=1}^{N}\Gamma_k e^{\text{i} D(t_k)}\right\vert^2,  \label{DP-N}
\ee
where $\Gamma_{k}=4$i$\lim\limits_{t\rightarrow t_{k}}(t-t_{k})\dot{\vartheta}(t)$; usually $\Gamma_{k}=1$ or $-1$.
Here $\dot{\vartheta} (t)$ accounts for the nonadiabatic coupling
between the adiabatic states, with $\vartheta(t)=\frac12 \tan^{-1}
\Omega(t)/\Delta(t)$.

\subsection{Optimization of two-state adiabatic passage}

Gu\'{e}rin \emph{et al.} \cite{Guerin} have used the DDP method to
optimize the adiabatic passage between two states in a very simple
manner. Assuming that the probability for nonadiabatic losses is
solely due to the transition points $t_{k}$, they have proposed to
suppress these altogether by choosing the Rabi frequency $\Omega
(t)$ and the detuning $\Delta (t)$ such that there are \emph{no
transition points}. This condition is obviously fulfilled if the
quasienergy splitting is \emph{constant}, \be
\varepsilon(t)=\sqrt{\Omega (t)^2+\Delta (t)^2}=\text{const.}
\label{trajectory} \ee For example, this condition is fulfilled
for a detuning and Rabi frequency defined as \bse
\begin{gather}
\Delta (t)=\varepsilon_0\cos \left[ \pi f(t)\right] ,\qquad \Omega (t)=\varepsilon_0\sin \left[ \pi f(t)\right] ,  \label{omega-delta} \\
0=f(-\infty )\leqq f(t)\leqq f(\infty )=1,  \label{f(t)}
\end{gather}
\ese
with $f(t)$ being an arbitrary monotonically increasing function
with the above property. Condition (\ref{trajectory}) is not the
only possible condition for adiabatic optimization, but it is the
simplest one \cite{Guerin}.

\section{Optimization of STIRAP\label{Sec-STIRAP}}

\subsection{STIRAP}

The probability amplitudes of the three states in STIRAP $\mathbf{c}(t)=\left[ c_1(t),c_2(t),c_3(t)\right]^{T}$ satisfy the Schr\"{o}dinger equation,
\be
\label{Sch-eq}
\text{i}\frac{\d}{\d t}\mathbf{c}(t) = \mathbf{H}(t)\mathbf{c}(t),
\ee
where the STIRAP Hamiltonian within the rotating-wave approximation (RWA) reads \cite{B.Shore}
\be
\mathbf{H}(t)=\tfrac12\left[ \begin{array}{ccc}
0 & \Omega_{p}(t) & 0 \\
\Omega_{p}(t) & 2\Delta  & \Omega_{s}(t) \\
0 & \Omega_{s}(t) & 0
\end{array}\right] .  \label{H}
\ee
The time-varying Rabi frequencies $\Omega_p(t)$ and
$\Omega_s(t)$ describe the couplings between the intermediate
state $\psi_2$ and, respectively, the initial state $\psi_1$ and
the target final state $\psi_3$. STIRAP is easily explained with
the so-called \emph{dark state} $\varphi_{d}(t)$, which is a
zero-eigenvalue eigenstate of $\mathbf{H}(t)$, \be
\label{dark-state} \varphi_{d}(t)=\psi_1\cos \vartheta
(t)-\psi_3\sin \vartheta (t), \ee where the time-dependent mixing
angle $\vartheta (t)$ is defined as \be
\vartheta(t)=\tan^{-1}\frac{\Omega_{p}(t)}{\Omega_{s}(t)}\text{.}
\label{theta} \ee The pulses in STIRAP are ordered
counterintuitively, i.e., the Stokes pulse precedes the pump
pulse, \be \underset{t\rightarrow -\infty
}{\lim}\frac{\Omega_{p}(t)}{\Omega_{s}(t)}=0,\qquad
\underset{t\rightarrow +\infty }{\lim
}\frac{\Omega_{s}(t)}{\Omega_{p}(t)}=0.\ \label{pulse-sequence}
\ee Then $0\overset{-\infty \leftarrow t}{\longleftarrow}\vartheta
(t)\overset{t\rightarrow\infty}{\longrightarrow}\pi/2$,
 and therefore the dark state $\varphi_d(t)$ connects adiabatically
states $\psi_1$ and $\psi_3$,
\be
\psi_1\overset{-\infty\leftarrow t}{\longleftarrow }\varphi_{d}(t)\overset{t\rightarrow\infty }{\longrightarrow }-\psi_3.
\ee
Thus, if the evolution is adiabatic then the population passes
from state $\psi_1$ to state $\psi_3$. Moreover, because the dark
state does not contain a contribution from the (possibly lossy)
intermediate state $\psi_2$, the properties of the latter are less
important.

When the evolution is not completely adiabatic, the population
transfer $\psi_1\rightarrow \psi_3$ might be incomplete. Moreover,
the intermediate state receives some transient population, which
may either be lost if state $\psi_2$ decays to other states, or
lead to decoherence if it decays back into $\psi_1$ or $\psi_3$.
For STIRAP to be a viable tool for quantum computing, nonadiabatic
transitions have to be reduced below the fault-tolerance limit of
$\lesssim 10^{-4}$ \cite{Nielsen00,fault tolerant QC}. We can
estimate the required resources for STIRAP to reach such a
fidelity as follows. The adiabatic condition for STIRAP (for
$\Delta =0$) demands large pulse areas $A_{p,s}=\int_{-\infty
}^{\infty }\Omega_{p,s}(t)$d$t\gg 1$. This global condition is
derived from the local adiabatic condition, which reads $\Omega
(t)\gg \left\vert\dot{\vartheta}(t)\right\vert $. The probability
for nonadiabatic transitions in the perturbative limit is
$P_{\text{na}}(t)\sim\dot{\vartheta}^2(t)/\Omega^2(t)$; it
measures the population that escapes to other adiabatic states and
reduces the fidelity. The infidelity is therefore $1-P_3\sim
1/A_{p,s}^2$. The fault-tolerance QIP limit therefore requires
$A_{p,s}\gtrsim 100$. In fact, this very rough estimate neglects
various details, such as the peculiarities of the nonadiabatic
coupling $\dot{\vartheta}(t)$, and in reality the condition for
the pulse areas is more restrictive.

In the following, we show that an optimized version of STIRAP can
reach the fault-tolerance QIP limit by using much smaller pulse
areas. In order to optimize STIRAP\ we use the same ideas as in
the two-state adiabatic passage optimization by Gu\'{e}rin
\emph{et al.} \cite{Guerin}. To this end we make use of the
reduction of STIRAP to effective two-state problems.

\subsection{Effective two-state systems\label{Sec-two-state}}

\subsubsection{Single-photon resonance}

On single-photon resonance ($\Delta =0$) the three-state system is
reduced to an effective two-state system, with a detuning
$\Omega_s(t)$\ and a coupling $\Omega_p(t)$
\cite{Laine,Vitanov96},
\be
\text{i}\frac{d}{dt}\left[
\begin{array}{c}
b_1(t) \\
b_2(t)
\end{array}\right]
 = \tfrac12\left[ \begin{array}{cc}
\Omega_s(t) & \Omega_p(t) \\
\Omega_p(t) & -\Omega_s(t)
\end{array}\right]
\left[ \begin{array}{c}
b_1(t) \\
b_2(t)
\end{array}\right] ,  \label{2-st.}
\ee
where the probability amplitudes $b_{1,2}(t)$ are related to $c_{1,2,3}(t)$ as follows \cite{Laine,Vitanov96}
\bse
\bea
c_1(t) &=& 2\text{Re} \left[ b_1^\ast (t)b_2(t)\right] \sin \vartheta (t) \notag\\
&& + \left( |b_1(t)|^2 - |b_2(t)|^2\right) \cos \vartheta (t), \label{diabatic-2st-vect} \\
c_2(t) &=& 2\i\text{Im}\left[ b_1^\ast (t) b_2(t)\right] , \\
c_3(t) &=& 2\text{Re} \left[ b_1^\ast (t) b_2(t)\right] \cos \vartheta (t) \notag\\
&&- \left( |b_1(t)|^2 - |b_2(t)|^2\right) \sin \vartheta (t).
\eea
\ese
Because we have for STIRAP $\vartheta (-\infty)=0$ and $\vartheta (\infty)=\pi/2$, the initial condition $c_1(-\infty)=1$
demands the condition $|b_1(-\infty)| =1$ in the effective two-state system.
The final-state population in STIRAP is
\be\label{c3-prob}
|c_3(+\infty)|^2 = \left[ |b_1(+\infty)|^2 - |b_2(+\infty)|^2\right]^2.
\ee
Consequently, an optimized adiabatic evolution in the two-state system \eqref{2-st.} implies optimized STIRAP.
Hence applied to STIRAP, the two-state optimum condition \eqref{trajectory} simply yields
\be\label{optimal resonance}
\Omega_p(t)^2 + \Omega_s(t)^2 = \Omega^2 = \text{const}.
\ee
In other words, the rms Rabi frequency should be constant.
Again, as in the two-state optimization, this is not the only possible
optimization condition but it is the simplest one.

\subsubsection{Far-off-resonance fields}

For large single-photon detuning $\Delta(t)$, the intermediate
state can be eliminated adiabatically by setting $\dot{c}_2(t)=0$
in Eq. (\ref{Sch-eq}). We thus obtain \cite{STIRAP-reviews}

\be
\i\frac{\d}{\d t} \left[ \begin{array}{c} c_1(t) \\ c_3(t) \end{array} \right]
 =\tfrac12 \left[ \begin{array}{cc} -\Delta_{\text{eff}}(t) & \Omega_{\text{eff}}(t) \\
\Omega_{\text{eff}}(t) & \Delta_{\text{eff}}(t) \end{array}\right] \left[ \begin{array}{c} c_1(t) \\ c_3(t) \end{array}\right] ,  \label{two-state-eff-large-d}
\ee
where the effective Rabi frequency $\Omega_{\text{eff}}(t)$ and detuning $\Delta_{\text{eff}}(t)$ are defined as
\bse
\bea
\Omega_{\text{eff}}(t) &=&-\frac{\Omega_{p}(t)\Omega_{s}(t)}{2\Delta (t)}, \label{Rabi-eff} \\
\Delta_{\text{eff}}(t)
&=&\frac{\Omega_{p}(t)^2-\Omega_{s}(t)^2}{4\Delta (t)}
\label{Detun-eff}
\eea
\ese
The two-state condition for optimal adiabatic passage now reads
\be \label{optimal far off resonance}
\Omega_{\text{eff}}(t)^2+\Delta_{\text{eff}}(t)^2 = \left[
\frac{\Omega_p(t)^2+\Omega_{s}(t)^2}{4\Delta (t)}\right]^2 =
\text{const}.
\ee

For constant $\Delta$ this condition is identical to the one we
found on resonance, Eq. (\ref{optimal resonance}), that is it
requires a constant rms Rabi frequency $\Omega $. We point out
that, due to the identical conditions, on and off single-photon
resonance, optimization is ensured over a very wide range of
single-photon detunings.

\section{Optimization of STIRAP: examples \label{Sec-implementation}}

\subsection{Pulse shapes}

Conditions (\ref{optimal resonance}) and (\ref{optimal far off
resonance}) suggest the following parameterization of the pump and
Stokes fields
\bse
\bea
\Omega_{p}(t) &=&\Omega \sin \left[ \frac{\pi }2f(t)\right] , \\
\Omega_{s}(t) &=&\Omega \cos \left[ \frac{\pi }2f(t)\right] ,
\eea
\ese
where $f(t)$ is an arbitrary monotonically increasing function,
$0=f(-\infty )\leqq f(t)\leqq f(\infty )=1$. Viewed
mathematically, $\Omega_{p}(t)$ and $\Omega_{s}(t)$ define an
adiabatic path, for which the nonadiabatic correction given by the DDP
formula vanishes, which leads to an optimal adiabatic following of
the dark state.

The exact fulfillment of condition (\ref{optimal resonance})
requires a constant $\Omega $ and hence, infinite pulse areas.
This unphysical condition can be overcome by using a ``mask''
function $F(t)$,
\bse
\bea
\Omega_{p}(t) &=&\Omega_0F(t)\sin \left[ \frac{\pi }2f(t)\right] , \\
\Omega_{s}(t) &=&\Omega_0F(t)\cos \left[ \frac{\pi }2f(t)\right] .
\eea
\ese
Then the rms Rabi frequency becomes \emph{time-dependent}, $\Omega
(t)=\Omega_0F(t)$. This replacement does not necessarily violate
the optimization condition (\ref{optimal resonance}) because DDP
transition points may still be absent, e.g. if the mask function
has a suitable shape, such as Gaussian. Still, a pulse-shaped mask
$F(t)$ violates the DDP conditions because the eigenenergy
splitting $\varepsilon(t)\equiv\Omega_0F(t)$ becomes degenerate as
$t\rightarrow \pm \infty $. The implication is that the
probability for nonadiabatic transitions as a function of the
pulse area is no longer expressed as a simple exponential,
Eq.~\eqref{DP-1}, but rather by a sum of an exponential and an
oscillatory term with an amplitude that vanishes only polynomially
with the pulse area \cite{Vitanov96,Drese98}. The exponential term
dominates for moderate pulse areas, whereas the oscillatory
polynomial term dominates for large areas
\cite{Vitanov96,Drese98}. The border value of the area $A_b$,
where the exponential decline breaks down into slowly damped
oscillations, is proportional to the ratio R between the rms pulse
area $A=\sqrt{A_p^2+A_s^2}$ and the overlap area $A_o$ (the area
of overlap of the pump and Stokes pulses) $A_b\propto R= A/A_o$.
The exponential decline of nonadiabatic transitions is favorable
for high-fidelity STIRAP because it allows one to achieve high
fidelity with moderate pulse areas. This is turn implies a large
value of the breakdown area $A_b$, so that the (slowly damped)
oscillations emerge only when the infidelity is very low. Hence
high-fidelity STIRAP is facilitated by asymmetric pulses, with
longer outer tails, so that the ratio $R$ (and hence the breakdown
area $A_b$) is large. These observations are further illustrated
with the figures below.

The above arguments suggest the following recipe for choosing $F(t)$:
\begin{itemize}
\item $F(t)$ should be a \emph{flat} function during the time of overlap of the pump and Stokes pulses, during which the population transition takes place;

\item $F(t)$ should have a sufficiently \emph{large width} so that the rms-to-overlap ratio $R$ is sufficiently large.
\end{itemize}

In the examples below, as a mask $F(t)$ we use the hypergaussian function,
\be
F(t)=\text{e}^{-\left( t/T_0\right)^{2n}},  \label{F-pulse}
\ee
where $n=1$ corresponds to the Gaussian shape.
For larger (positive integer) $n$ the condition $F(t)\simeq\,\text{const}$ is fulfilled increasingly better in the overlap region.

There is a some leeway in the choice of the function $f(t)$ as
long as the adiabatic condition is fulfilled in the overlap
region. We use
 \be
 f(t) = \dfrac1{1+\text{e}^{-\lambda t/T}}.  \label{f(t) example}
\ee

\begin{figure}[tb]
\includegraphics[width=75mm]{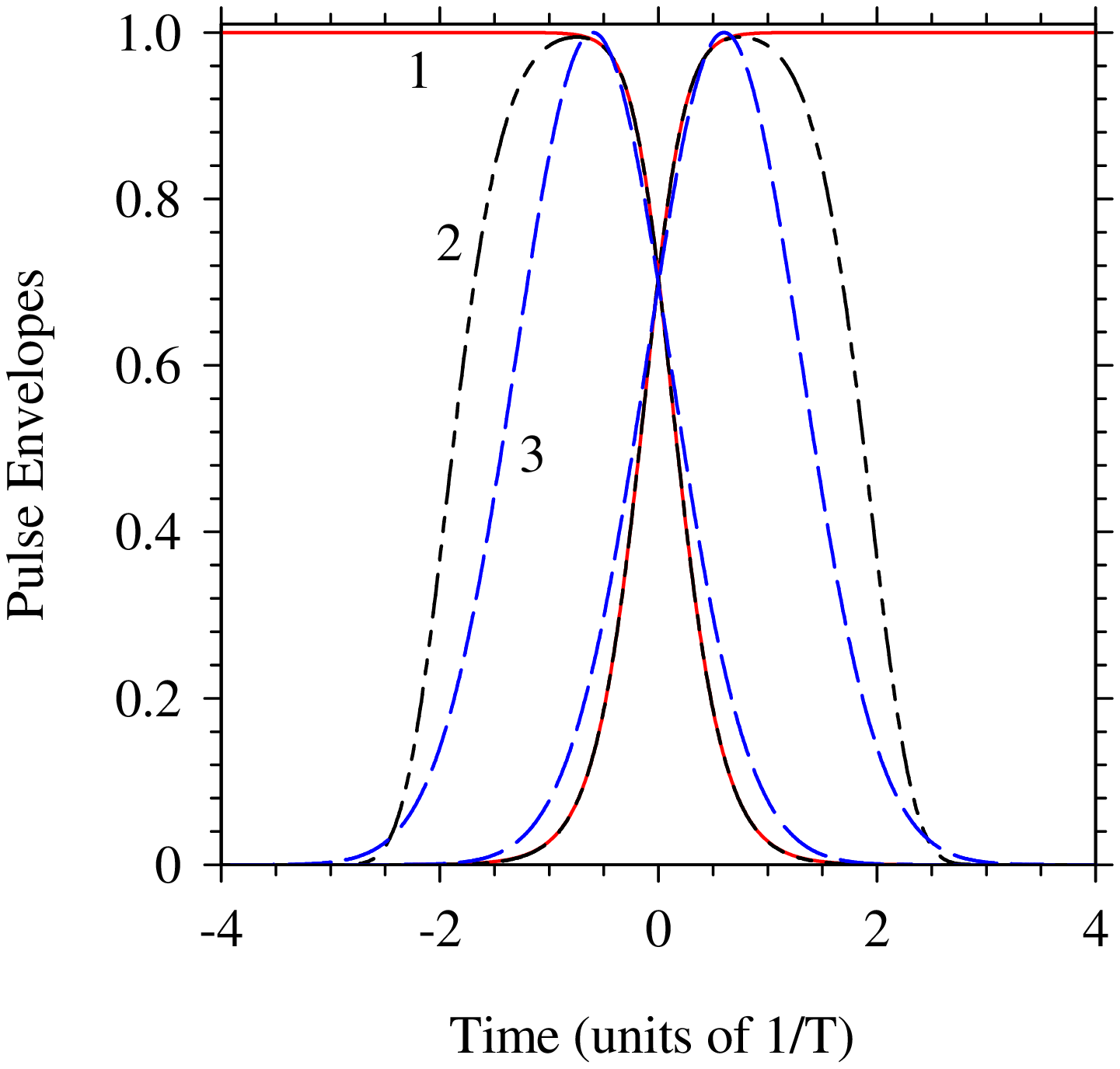}
\caption{Pulse shapes.
 1: ideally optimized pulse shapes, which obey the optimization condition (\protect\ref{optimal resonance}) for all times $t$;
 2: pulse shapes with a hypergaussian mask (\protect\ref{F-pulse}), with $n=3$, $\lambda=4$ and $T_0=2T$;
 3: Gaussian pulses (\protect\ref{Gaussian-pulses}), with pulse delay $\tau=1.2T$.}
\label{Fig-epspulses}
\end{figure}

We point out that other choices of the function $f(t)$ are also
possible. However, with the chosen method of optimization being
based on the DDP approximation, which is valid only in the
near-adiabatic regime, the function $f(t)$ has to fulfill the
adiabaticity criterion \be \label{2st. ad. condition}
 |\dot{\vartheta}(t)| \ll \ | \varepsilon (t) |,
\ee
where $\dot{\vartheta}(t)$ is the nonadiabatic coupling. Using
Eq. (\ref{2st. ad. condition}) we obtain the following condition
for the function $f(t)$
\be
 |\dot{f}(t)| \ll \Omega_0 | F(t) |;
\ee
hence $f(t)$ should have a smooth time dependence in
order to facilitate adiabaticity. Once in the adiabatic regime,
however, the function $f(t)$ does not affect the DDP optimization
because it does not appear in the condition \eqref{optimal
resonance}.

Three different pulse shapes are shown in Fig.
\ref{Fig-epspulses}: pulse shapes that obey the optimization
condition \eqref{optimal resonance} for all times $t$, along with
the more realistic example with a hypergaussian mask
\eqref{F-pulse}, that obey the optimization condition
\eqref{optimal resonance} only in the region of overlap of the
pump and Stokes pulses, and Gaussian pulses.
\be
\label{Gaussian-pulses}\Omega_p =\Omega_0\e^{-(t-
\tau/2)/T^2},\qquad \Omega_s = \Omega_0 \e^{-(t+\tau/2)/T^2},
\ee
where $\tau $ is the pulse delay. We note that the pulse area $A$
for the Gaussian pulses is almost identical with the DDP optimized
pulses. In the following we demonstrate that the optimally shaped
pulses are superior to the Gaussian pulses, even with optimized
delay for the latter, in achieving a very high fidelity.

\subsection{Examples of ultrahigh-fidelity STIRAP}

\begin{figure}[tb]
\includegraphics[width=75mm]{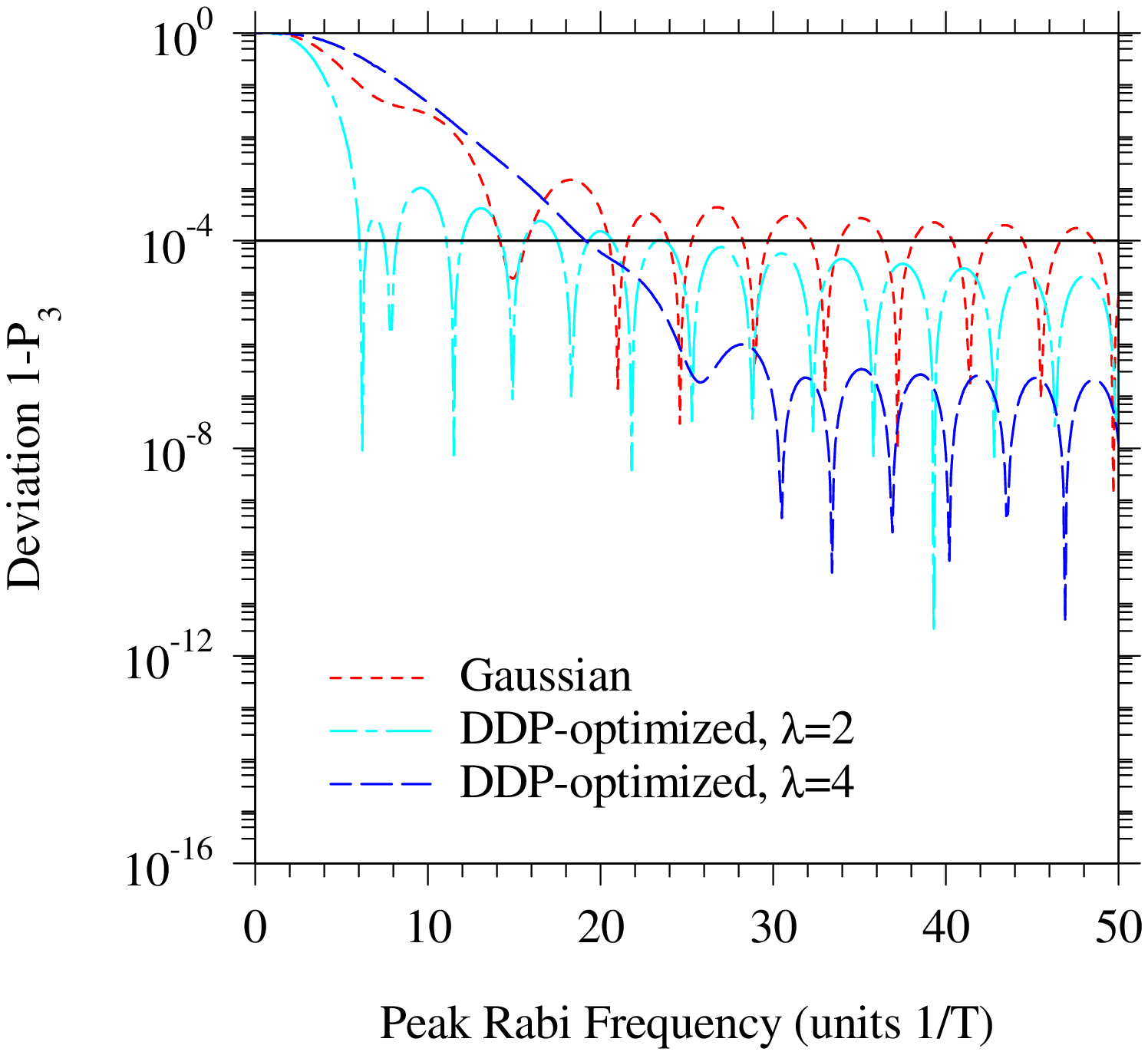}
\caption{Deviation (infidelity) from complete population transfer
vs the peak Rabi frequency for Gaussian (delay $\tau=1.2T$) and
DDP-optimized pulses \eqref{F-pulse} ($n=3$, $\lambda=4$, $T_0=2T$
and $n=1$, $\lambda=2$, $T_0=2T$).} \label{Fig-epsopt_exp1}
\end{figure}

In Figure \ref{Fig-epsopt_exp1} we plot the STIRAP infidelity,
i.e. the deviation $1-P_3$ from perfect transfer for the optimized
pulses, described above, and compare these to the results for the
traditional implementation of STIRAP with a pair of Gaussian
pulses. The infidelity is shown as a function of the peak Rabi
frequency. For Gaussian pulses, the pulse delay is chosen such
that a nearly maximum fidelity is obtained. Despite the optimum
delay, the Gaussian shapes do not allow us to reduce the
infidelity below the limit $10^{-4}$ in the shown range of pulse
areas (eventually, at very high pulse areas the infidelity drops
below this limit). On the contrary, DDP-optimized pulse shapes
(with $n=1$, $\lambda=2$, $T_0=2T$) break this limit even for
small pulse areas.

\begin{figure}[tb]
\includegraphics[width=75mm]{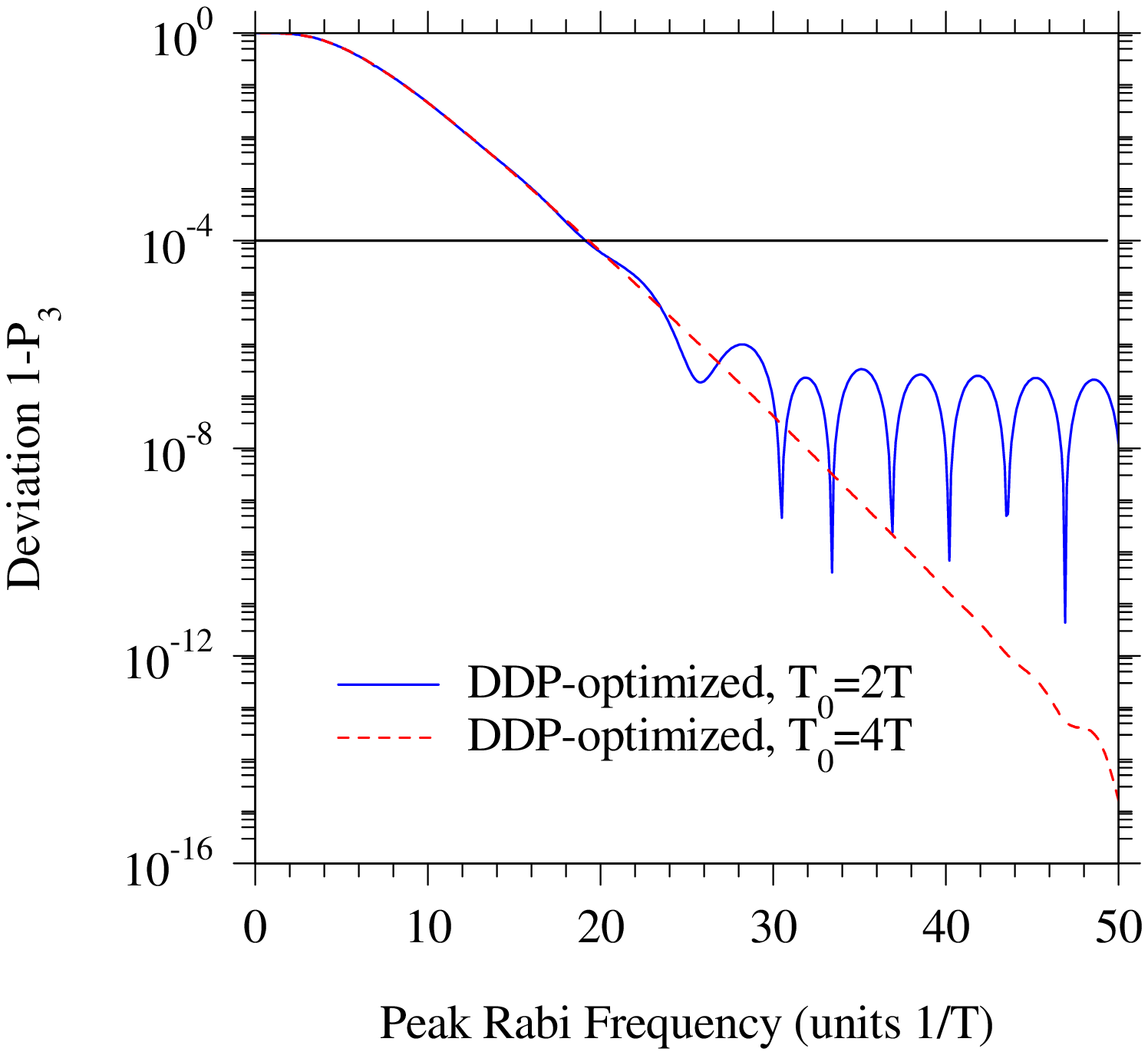}
\caption{Deviation (infidelity) from complete population transfer
vs the peak Rabi frequency for DDP optimized pulses for different
values of the width $T_0$ of the hypergaussian ($n=3$) mask
$F(t)$, Eq. (\protect\ref{F-pulse}).} \label{Fig-epsopt}
\end{figure}

We note that the DDP pulses are not fully optimized due to the
mask function $F(t)$ vanishing at large times. Therefore the
respective fidelity curve starts to oscillate, which signals the
occurence of nonadiabatic transitions. However, the magnitude of
these nonadiabatic transitions can be controlled by the width of
the mask function $F(t)$: a larger width pushes these oscillations
further down. Hence even for small pulse areas, the optimally
shaped pulses are far superior to the Gaussian pulses, as shown in
Fig. \ref{Fig-epsopt_exp1}. This tendency is also visible in Fig.
\ref{Fig-epsopt} where the infidelity is plotted as a function of
the pulse area for two different widths $T_0$ of the mask function
$F(t)$, Eq. (\ref{F-pulse}). By increasing the mask width $T_0$,
the validity range of the adiabatic optimization condition
(\ref{optimal resonance}) widens, and the optimized pulses
approach the ideal DDP-optimized pulse shapes in Fig.
\ref{Fig-epspulses}.

\begin{figure}[tb]
\includegraphics[width=75mm]{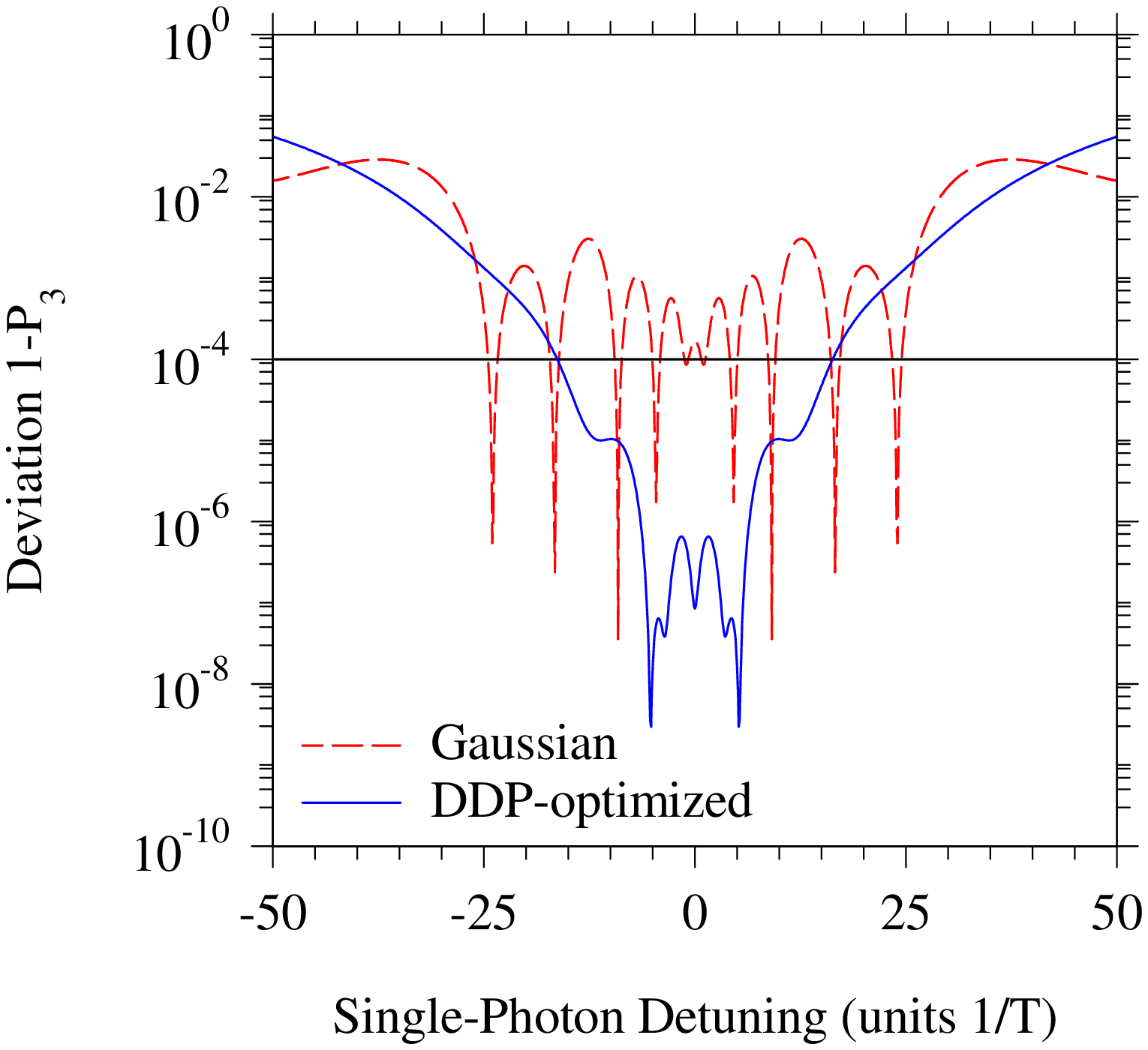}
\caption{Line profile as a function of the common detuning
$\Delta$ from the intermediate level for DDP-optimized ($n=3$,
$\lambda=4$, $T_0=2T$) and Gaussian pulses (delay $\tau =1.2T$)
for a peak Rabi frequency $\Omega=20$.} \label{Fig-epsdetuning}
\end{figure}

In Fig.~\ref{Fig-epsdetuning} we compare the line profile as a
function of the common detuning $\Delta$ from the intermediate
level (see Eq. (\ref{H}))  for the optimized and Gaussian pulse
shapes. The time delay for the Gaussian pulses is numerically
chosen for nearly maximum fidelity. It is known that a
single-photon detuning does not affect the dark state (as long as
two-photon resonance is maintained) \cite{STIRAP-reviews}.
Nonetheless, adiabaticity deteriorates and the transfer efficiency
decreases with increasing single-photon detuning. The robustness
of the high fidelity STIRAP (i.e. STIRAP where the infidelity is
below the limit $10^{-4}$) against the single-photon detuning is
much more pronounced for the optimized pulses. This feature is
readily explained by the fact that the same pulse shapes optimize
STIRAP both for $\Delta =0$ and for large $\Delta $, as discussed
above.

\begin{figure}[tb]
\includegraphics[width=75mm]{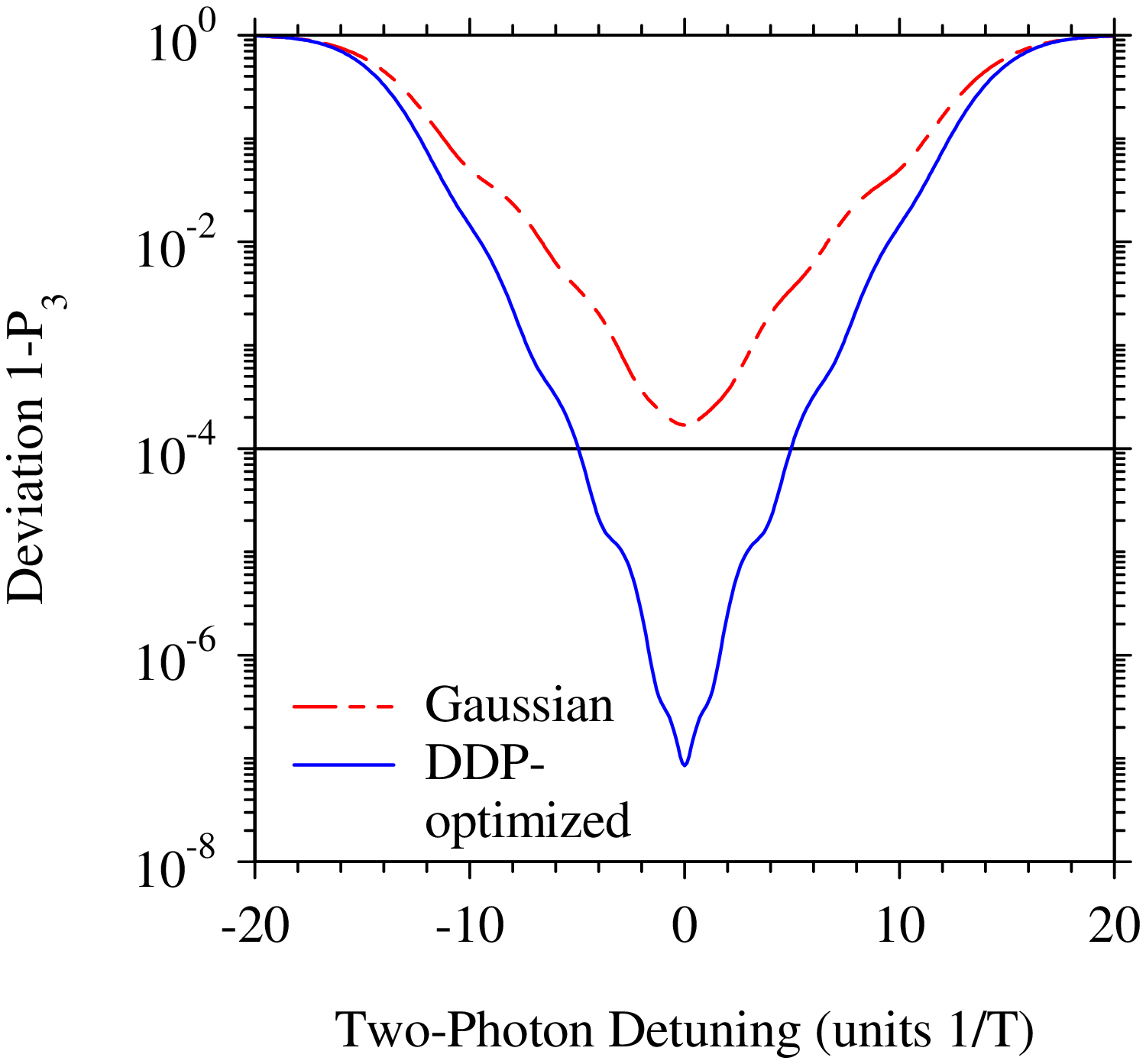}
\caption{Raman line profile for the DDP-optimized ($n=3$,
$\lambda=4$ $T_0=2T$) and Gaussian pulses (delay $\tau =1.2$) for
a peak Rabi frequency $\Omega=20$.} \label{Fig-eps2pdet}
\end{figure}

It is known that the transfer efficiency for STIRAP is much more
sensitive to a detuning from Raman resonance than to the
single-photon detuning \cite{STIRAP-reviews}. Figure
\ref{Fig-eps2pdet} shows the Raman line profile for DDP-optimized
and Gaussian pulses. As before, the time delay for the Gaussian
pulses is numerically chosen such that a nearly maximum transition
probability $P_3$ is obtained. The optimized pulse shapes are far
superior to the Gaussian pulses and allow one to maintain a high
fidelity over a wide range of two-photon detunings.

\subsection{Relative error due to the RWA approximation\label%
{RWA}}

The rotating wave approximation (RWA) is widely used whenever
laser-induced excitations with optical frequencies $\omega$ much
larger than the Rabi frequency $\Omega$ are considered. Typical
optical (carrier) frequencies  are $\omega\sim 10^{16} [s^{-1}]$,
while the typical Rabi frequencies are within the range
$\Omega\sim 10^{8}-10^{9} [s^{-1}]$ \cite{B.Shore}. For two-level
systems, perturbative inclusion of the counter-rotating terms
results in the Bloch-Siegert shift of the eigenenergies by $\sim
\Omega^{2}/\omega$ \cite{Silverman}. Within a typical pulse
duration $T \sim 10^{-6}-10^{-8} [s]$, this accumulates maximum
deviation from the calculated probabilities of
\begin{equation}
\triangle P_{e} \sim (T \frac{\Omega^{2}}{\omega})^{2}\simeq
10^{-8}-10^{-16}
\end{equation}
It should be also noted that the results in RWA are exact in case
of circular polarization \cite{B.Shore}.

\section{Optimization of \red Hadamard gates \black \label{Sec-f-STIRAP}}

Fractional STIRAP (f-STIRAP) is a variation of STIRAP, which
creates an arbitrary preselected coherent superposition of states
$\psi_1$ and $\psi_3$,
\be\label{Hadamard}
\Psi
=\psi_1\cos\alpha-\psi_3\sin\alpha.
\ee
As in STIRAP, the Stokes pulse precedes the pump pulse, but unlike
STIRAP, where the Stokes pulse vanishes first, in f-STIRAP the two
pulses vanish simultaneously, while maintaining a constant ratio
of their amplitudes \cite{Marte},
\be \label{pulse-sequence-fSTIRAP} 0\overset{-\infty \leftarrow
t}{\longleftarrow
}\frac{\Omega_{p}(t)}{\Omega_{s}(t)}\overset{t\rightarrow \infty
}{\longrightarrow }\tan \alpha .
\ee
A convenient realization of
f-STIRAP reads \cite{fSTIRAP1}
\bse\label{f-STIRAP}
\bea
\Omega_{p}(t) &=&\Omega_0\text{e}^{-\left( t-\tau /2\right)^2/T^2}\sin \alpha ,  \label{fSTIRAP-pump} \\
\Omega_{s}(t) &=&\Omega_0\left\{ \text{e}^{-\left( t+\tau /2\right)^2/T^2}+\text{e}^{-\left( t-\tau /2\right)^2/T^2}\cos \alpha\right\} .  \label{fSTIRAP-Stokes}
\eea
\ese
The DDP-optimized pulses, in analogy with the full STIRAP, read
\bse\label{optimal f-STIRAP}
\bea
\Omega_{p}(t) &=&\Omega_0F(t)\sin \left[ \alpha f(t)\right] , \\
\Omega_{s}(t) &=&\Omega_0F(t)\cos \left[ \alpha f(t)\right] ,
\eea
\ese
where $f(t)$ is again an arbitrary function that satisfies
condition \eqref{f(t)}. It is easy to verify that these pulse
shapes satisfy the f-STIRAP condition
(\ref{pulse-sequence-fSTIRAP}). For half-STIRAP, when an equal
coherent superposition of states $\psi_1$ and $\psi_3$ is created,
we should have $\alpha =\pi /4$. This superposition corresponds to
the Hadamard gate, which is one of the fundamental gates in
quantum information processing.

\begin{figure}[tb]
\includegraphics[width=75mm]{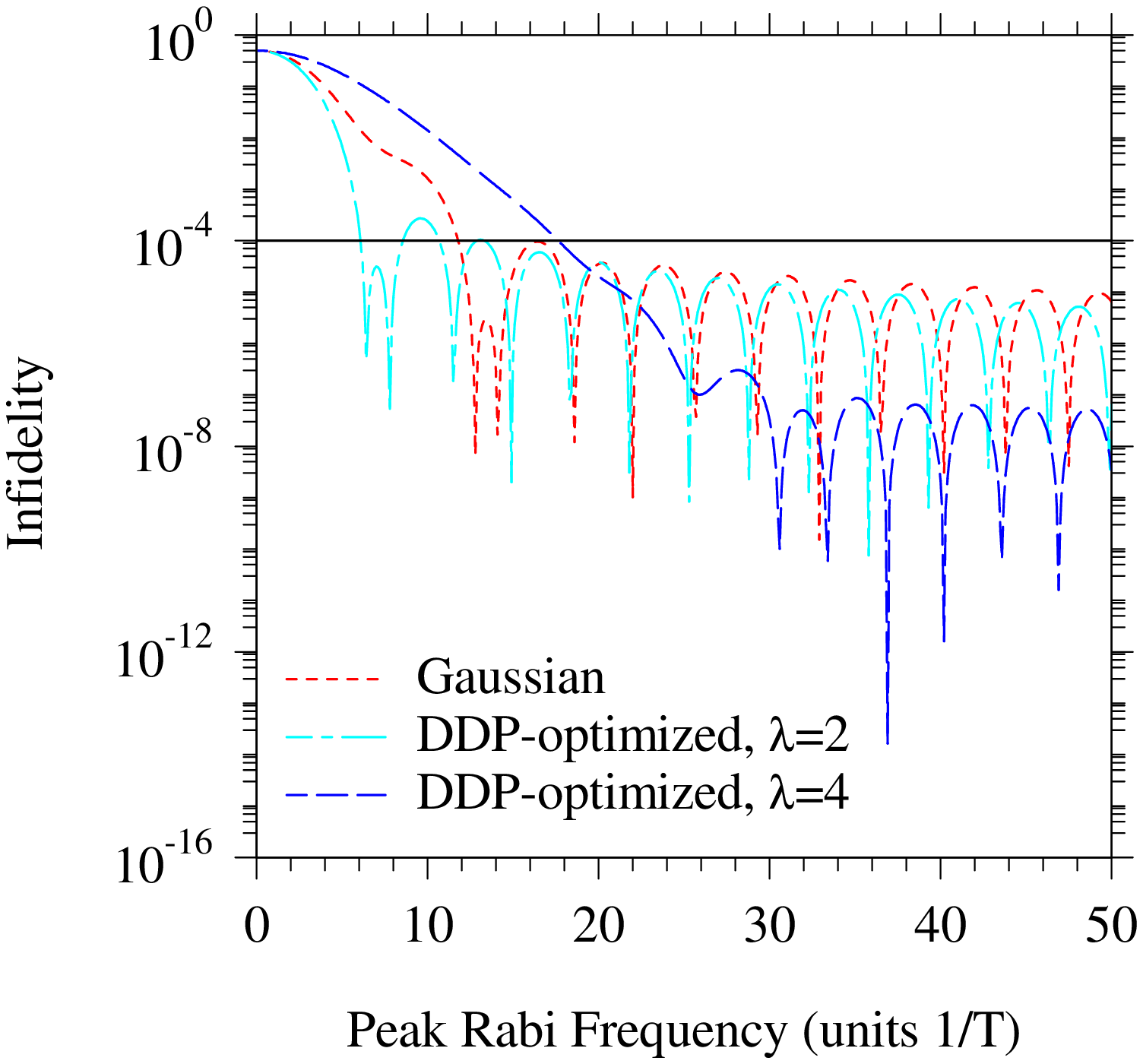}
\caption{Transfer efficiency for fractional STIRAP vs the peak
Rabi frequency for DDP-optimized pulses (with $n=3$, $\lambda=4$,
$T_0=2T$ and $n=1$, $\lambda=2$, $T_0=2T$) and Gaussian pulses
\eqref{f-STIRAP} (with $\tau =1.4T$).} \label{Fig-fstirap}
\end{figure}

Figure \ref{Fig-fstirap} compares the transfer efficiency of
f-STIRAP for DDP-optimized pulses and for pulses given by Eqs.
(\ref{f-STIRAP}). The pulse delay for the pulses (\ref{f-STIRAP})
is chosen numerically for nearly maximum fidelity. In Fig.
\ref{Fig-fstirap} the infidelity is defined as
\be\label{infidelity}
 1-|<\Psi_{final}|\Psi_{desired}>|^2,
\ee where $\Psi_{desired}$ is given by Eq. (\ref{Hadamard}) at
$\alpha =\pi /4$ and $\Psi_{final}$ is numerically calculated for
both DDP-optimized and Gaussian pulses. It is important to note
that despite the delay optimization, the pulses (\ref{f-STIRAP})
do not allow reduction of the infidelity below a certain limit.
Due to the mask function $F(t)$, the DDP-optimized pulses led to
an oscillatory behaviour of the infidelity. However, the
oscillatory regime and hence the fidelity depend on the width
$T_0$ and can be controlled. In particular in the regime of small
pulse areas, the DDP-optimized version of f-STIRAP (with $n=1$,
$\lambda=2$, $T_0=2T$) is far superior to the traditional Gaussian
pulses as shown in Fig. \ref{Fig-fstirap}.

\section{Conclusions\label{Sec-conclusion}}

We have proposed an optimization of the fidelity of the STIRAP
technique, which uses the DDP approach to minimize nonadiabatic
losses. The rationale for this is the reduction of STIRAP from
three to two states either: on exact single-photon resonance or
for large single-photon detuning. Interestingly, the optimized
pulse shapes are the same in both regimes, which makes
DDP-optimized STIRAP very robust against variations in the
detuning. We have demonstrated with numerical simulations that the
fidelity of this optimized STIRAP can reach very high values, with
an error well below the fault-tolerant QIP limit of $10^{-4}$,
which is very difficult to reach by optimizing the pulse delay
with the usual Gaussian pulse shapes (or other symmetric pulse
shapes, such as hyperbolic-secant). The proposed optimization is
of potential importance for QIP because it supplements the
robustness of STIRAP against parameter variations with an
ultrahigh fidelity of gate operations. We furthermore emphasize
that a similar optimization method could also have a significant
impact on the vacuum-stimulated Raman scattering in cavity QED
\cite{Kuhn02-07}.

\acknowledgments This work has been supported by the projects
EMALI, FASTQUAST and SCALA of the European Union, the EPSRC
grant(EP/E023568/1), the research unit 635 of the German research
foundation and the Bulgarian NSF Grants Nos. 205/06, 301/07, and
D002-90/08.

\end{document}